\documentclass[11pt]{article}
\usepackage{moriond,epsfig}
\usepackage{amsmath}
\usepackage{subfigure}

\def\be{\begin{equation}}
\def\ee{\end{equation}}
\def\bea{\begin{eqnarray}}
\def\eea{\end{eqnarray}}

\begin{document}
\vspace*{4cm}
\title{Central Exclusive Meson Pair Production in the Perturbative Regime}

\author{\footnote{KRYSTHAL collaboration} L.A. Harland-Lang \footnote{speaker}}
\address{Cavendish Laboratory, University of Cambridge, J.J.\ Thomson Avenue, Cambridge, CB3 0HE, UK}
\author{V.A. Khoze}
\address{Department of Physics and Institute for Particle Physics Phenomenology, University of Durham, DH1 3LE, UK}
\author{M.G. Ryskin}
\address{Petersburg Nuclear Physics Institute, Gatchina, St. Petersburg, 188300, Russia}
\author{W.J. Stirling}
\address{Cavendish Laboratory, University of Cambridge, J.J.\ Thomson Avenue, Cambridge, CB3 0HE, UK}

\maketitle\abstracts{We present a study of the central exclusive production (CEP) of meson pairs~\cite{HarlandLang:2011qd}, $M\overline{M}$, at sufficiently high invariant mass that a perturbative QCD formalism is applicable. Within this framework, $M\overline{M}$ production proceeds via the $gg\to M\overline M$ hard scattering sub-process, which can be calculated within the hard exclusive formalism. We present explicit calculations for the $gg \to M\overline M$ helicity amplitudes for different meson states and, using these, show results for meson pair CEP in the perturbative regime.}
%
%
%
\noindent Central exclusive production processes of the type
\begin{equation}\label{exc}
pp({\bar p}) \to p+X+p({\bar p})\;,
\end{equation}
can significantly extend the physics programme at high energy hadron colliders~\cite{KMRprosp,fp420,Albrow:2010zz}. Here $X$ represents a system 
of invariant mass $M_X$, and the `$+$' signs denote the presence of large rapidity gaps. Such reactions provide a very promising way to investigate both QCD dynamics and new physics in hadron collisions~\cite{Khoze04,HarlandLang09,HarlandLang10,khrysthal,HKRSTW,HKRSTW1,shuvaev,Albrow:2010yb}, providing an especially clean environment in which to measure the nature and quantum numbers (in particular, the spin and parity) of new states.

These processes have been measured at the Tevatron by the CDF collaboration, who  have published a search for $\gamma \gamma$ CEP~\cite{cdf:2007na} with $E_T(\gamma) >$ 5 GeV, and many more candidate events have been observed \cite{Albrow:2010zz} by lowering the $E_T(\gamma)$ threshold to $\sim$2.5 GeV. This process (together with charmonium CEP, the observation of which was reported in~\cite{Aaltonen:2009kg}), can serve as a `standard candle' reaction with which we can check the predictions for new physics CEP at the LHC \cite{HarlandLang10,Khoze:2004ak}. A good quantitative theoretical understanding of the  $\pi^0\pi^0$ CEP background is therefore crucial, as one or both of the 
photons from $\pi^0 \to \gamma\gamma$ decay can mimic the `prompt' photons from $gg \to \gamma\gamma$ CEP.
 
As discussed in~\cite{Khoze04,HarlandLang09,HarlandLang10}, the observation of $\chi_{c0}$ CEP via two-body decay channels to light mesons
is of special interest for both studying the dynamics of heavy quarkonia and for testing the QCD framework of CEP. However, in this case we may  
 expect a sizeable background resulting from direct QCD meson pair production; such a non-resonant contribution should therefore be carefully evaluated.

Studies of meson pair CEP would also present a new test of the perturbative formalism, with all 
its non-trivial ingredients, from the structure of the hard sub-processes to the incorporation of rescattering effects 
of the participating particles. Recall~\cite{Khoze00a} that in exclusive processes, the incoming $gg$ state satisfies 
special selection rules in the limit of forward outgoing protons, namely it has $J_z = 0$, 
where $J_z$ is the projection of the total $gg$ angular momentum on the beam axis, and positive $C$ and $P$ parity.  
Hence only a subset of the helicity amplitudes for the $gg \to X$ sub-process contributes. The CEP mechanism therefore provides a 
unique possibility to test the polarization structure of the $gg \to X$ reaction.

\begin{figure}[h]
\begin{center}
\includegraphics[scale=0.5]{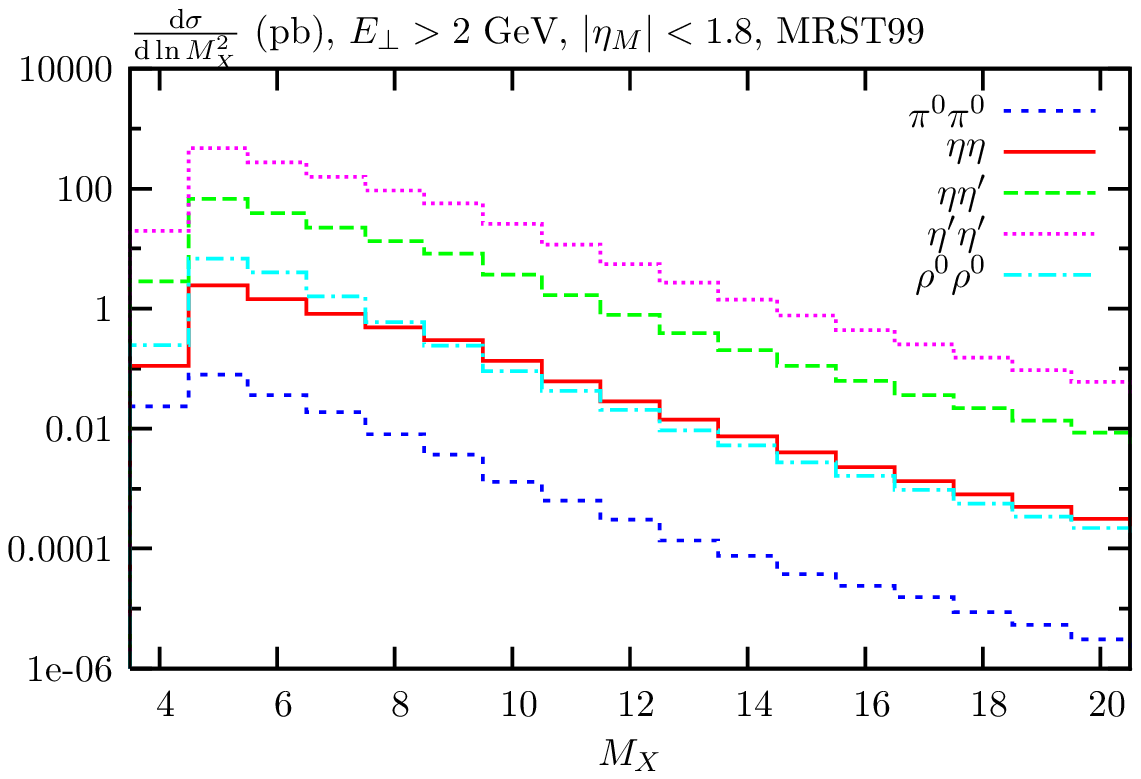}\qquad\qquad
\includegraphics[scale=0.5]{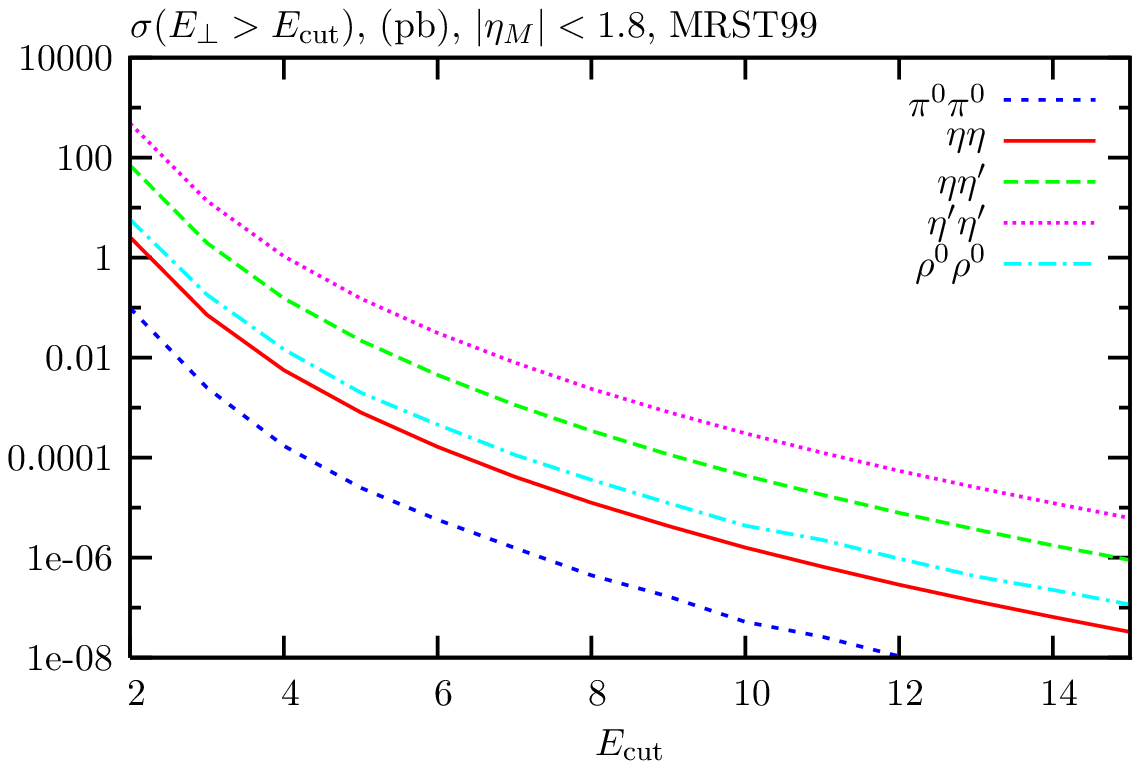}
\caption{${\rm d}\sigma/{\rm d}\ln M_X^2$ for meson transverse energy $E_\perp>2$ GeV, and cross section as a function of the cut $E_{\rm cut}$ on the meson $E_\perp$ at $\sqrt{s}=1.96$ TeV for the CEP of meson pairs, calculated within the perturbative framework.}\label{vsm}
\end{center}
\end{figure}

We will consider the $gg\to M\overline{M}$ process relevant to CEP within the `hard exclusive' formalism, which was used previously~\cite{Brodsky81,Benayoun89} to calculate the related $\gamma\gamma \to M\overline{M}$ amplitudes. The leading order contributions to $gg \to M\overline{M}$ can be written in the form
\begin{equation}\label{amp}
\mathcal{M}_{\lambda\lambda'}(\hat{s},\theta)=\int_{0}^{1} \,{\rm d}x \,{\rm d}y\, \phi_M(x)\phi_{\overline{M}}(y)\, T_{\lambda\lambda'}(x,y;\hat{s},\theta)\;.
\end{equation}
where $\hat{s}$ is the $M\overline{M}$ invariant mass, $\lambda$, $\lambda'$ are the gluon helicities and $\theta$ is the scattering angle in the $gg$ cms frame. $T_{\lambda\lambda'}$ is the hard scattering amplitude for the parton level process $gg\to q\overline{q}\,q\overline{q}$, where each (massless) $q\overline{q}$ pair is collinear and has the appropriate colour, spin, and flavour content projected out to form the parent meson. $\phi(x)$ is the meson wavefunction, representing the probability amplitude of finding a valence parton in the meson carrying a longitudinal momentum fraction $x$ of the meson's momentum.
\begin{figure}
\begin{center}
\subfigure[]{\includegraphics[scale=1.0]{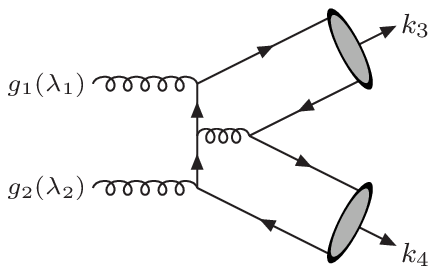}}\qquad\qquad
\subfigure[]{\includegraphics[clip,trim=-20 0 0 0,scale=0.6]{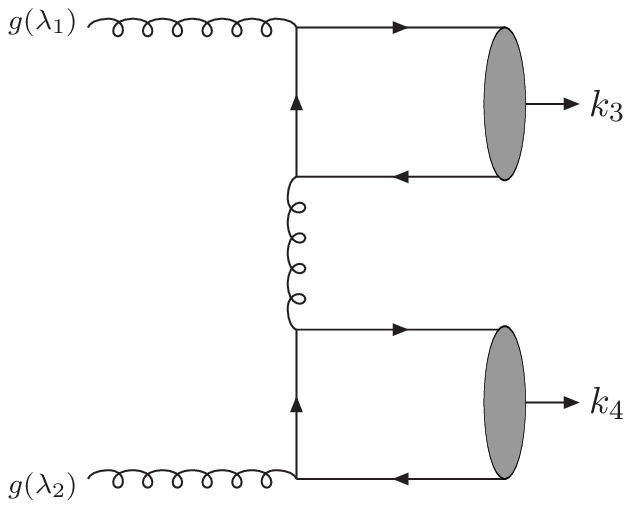}}
\caption{(a) A typical diagram for the $gg \to M\overline{M}$ process. (b) Representative `ladder' diagram, which contributes to the production of flavour-singlet mesons.}\label{ladder}
\end{center}
\end{figure}
We can then calculate the relevant parton-level helicity amplitudes for the $gg \to M\overline{M}$ process, for the production of scalar flavour-nonsinglet meson states ($\pi\pi$, $K^+K^-$, $K^0\overline{K}^0$). There are seven independent Feynman diagrams to compute-- a representative diagram is given in Fig.~\ref{ladder} (a). An explicit calculation gives
\begin{align}\label{T++}
T_{gg}^{++}=T_{gg}^{--}&=0\;,\\ \label{T+-}
T_{gg}^{+-}=T_{gg}^{-+}&=\frac{\delta^{\rm AB}}{N_C}\frac{64\pi^2\alpha_S^2}{\hat{s}xy(1-x)(1-y)}\frac{(x(1-x)+y(1-y))}{a^2-b^2\cos^2{\theta}}\frac{N_C}{2}\bigg(\cos^2{\theta}-\frac{2 C_F}{N_C}a\bigg)\;,
\end{align}
where $A,B$ are colour indices and
\begin{align}\label{a}
&a=(1-x)(1-y)+xy  &b=(1-x)(1-y)-xy\; .
\end{align}
We can see that the $gg\to M\overline{M}$ amplitude for $J_z=0$ gluons (\ref{T++}) vanishes at LO for scalar flavour-nonsinglet mesons, which, recalling the $J_z=0$ selection rule that strongly suppresses the CEP of non-$J_z=0$ states, will lead to a strong suppression (by $\sim$ two orders of magnitude) in the CEP cross section. However it should be noted that any NNLO corrections or higher twist effects which allow a $J_z=0$ contribution may cause the precise value of the cross section to be somewhat larger than the leading-order, leading-twist $|J_z|=2$ estimate, although qualitatively the strong suppression will remain. An important consequence of this is that the $\pi^0\pi^0$ QCD background to the $\gamma\gamma$ CEP process described above is predicted to be small. Some sample cross section plots for $\pi\pi$ CEP and the production of other meson states are shown in Fig.~\ref{vsm}.

We can also see that the $|J_z|=2$ amplitude (\ref{T+-}) vanishes for a particular value of $\cos^2\theta$. This vanishing of a Born amplitude for the radiation of massless gauge bosons, for a certain configuration of the final state particles is a known effect, usually labelled a `radiation zero'~\cite{Brodsky0s82}. The position of the zero is determined by an interplay of both the internal (in the present case, colour) and space-time (the particle $4$-momenta) variables, as can be seen in (\ref{T+-}), where the position of the zero depends on the choice of meson wavefunction, $\phi(x)$, through the variables $a$ and $b$, as well as on the QCD colour factors. However, it should again be noted that, as the $|J_z|=2$ amplitude is strongly suppressed by the $J_z=0$ selection rule, any NNLO or higher twist effects which allow a $J_z=0$ component to the cross section may give comparable contributions; it is therefore not clear that such a zero would in this case be seen clearly in the data. On the other hand, the destructive interference effects which lead to the zero in the $|J_z|=2$ amplitude (\ref{T+-}) will tend to suppress the CEP rate.

It is also possible for the $q\overline{q}$ forming the mesons to be connected by a quark line, via the process shown in Fig. \ref{ladder} (b). These amplitudes will only give a non-zero contribution for the production of ${\rm SU}(3)_F$ flavour-singlet states, i.e. $\eta'\eta'$ and, through $\eta$--$\eta'$ mixing, $\eta\eta$ and $\eta\eta'$ production. The relevant amplitudes are given by
\begin{align}\label{lad0}
T_{++}^{\rm lad.}=T_{--}^{\rm lad.}&=\frac{\delta^{AB}}{N_C}\frac{64\pi^2 \alpha_S^2}{\hat{s}xy(1-x)(1-y)}\frac{(1+\cos^2 \theta)}{(1-\cos^2 \theta)^2}\;,\\ \label{lad2}
T_{+-}^{\rm lad.}=T_{-+}^{\rm lad.}&=\frac{\delta^{AB}}{N_C}\frac{64\pi^2 \alpha_S^2}{\hat{s}xy(1-x)(1-y)}\frac{(1+3\cos^2 \theta)}{2(1-\cos^2 \theta)^2}
\end{align}
for the production of scalar mesons. As the $J_z=0$ amplitudes do not vanish, we will expect $\eta'\eta'$ CEP to be strongly enhanced relative to, for example, $\pi\pi$ production, due to the $J_z=0$ selection rule which operates for CEP. In the case of $\eta\eta$ production, the flavour singlet contribution will be suppressed by a factor $\sin^4 \theta_P \sim 1/200$, where $\theta_P$ is the octet-singlet mixing angle~\cite{Gilman:1987ax}, which may therefore be comparable to the $|J_z|=2$ flavour-octet contribution. In fact, after an explicit calculation we find that the $\eta\eta$ CEP cross section is expected, in the regions of phase space where the perturbative formalism is applicable,  to be dominant over $\pi\pi$ CEP. 

A further interesting possibility we should in general consider is a two-gluon Fock component $|gg\rangle$ to the $\eta(\eta')$ mesons, which can readily be included using the formalism outlined above. In particular, the relevant perturbative $gg\to 4g$ amplitude can be calculated in the usual way and, as this does not vanish for $J_z=0$ initial-state gluons, we may expect it to enhance the $\eta\eta$ and $\eta'\eta'$ CEP rates. This will depend sensitively on the size of the two-gluon wavefunction: therefore, by considering the CEP of $\eta(\eta')$ pairs at sufficiently high invariant mass, it may be possible to extract some information about the relative importance of the leading-twist quark and gluon wavefunctions.

Finally we have also calculated in the same way the amplitudes $T^{gg}_{\lambda_1\lambda_2,\lambda_3\lambda_4}$ for the $g(\lambda_1)g(\lambda_2)\to V(\lambda_3)\overline{V}(\lambda_4)$ process, where $V(\overline{V})$ are spin-1 mesons; for the sake of brevity, these are given explicitly elsewhere~\cite{HarlandLang:2011qd}. We should also in general consider a `non-perturbative' double-Pomeron-exchange picture for low values of the meson pair invariant mass, where we may not expect the perturbative framework described above to be applicable, see~\cite{HarlandLang:2011qd} for some discussion of this, as well as of a secondary perturbative mechanism, where both the $t$-channel gluons exchanged in the standard CEP picture couple to quark lines. We find that this process, which represents the perturbative tail of the non-perturbative production mechanism, is a power correction to the standard CEP process, and will therefore be strongly suppressed at high values of the meson pair $k_\perp$. We also note that the above $gg \to M\overline{M}$ helicity amplitudes can be considered within the MHV formalism~\cite{Mangano90}, which in some cases greatly simplifies the calculation -- see~\cite{HarlandLang:2011qd} for details.

To conclude, we have presented a study of the CEP of meson pairs in the perturbative regime, with the $gg\to M\overline{M}$ subprocess helicity amplitudes calculated within the hard exclusive formalism. This is of relevance as a background to $\gamma\gamma$ CEP in the case of $\pi^0\pi^0$ production, and to the CEP of heavy resonant states which decay to light meson pairs. Moreover, it is also a process which is important in its own right, allowing novel tests of the overall perturbative formalism as well as displaying various interesting theoretical properties.

LHL thanks the Organisers for providing an excellent scientific  environment at the Workshop.


\section*{References}
\begin{thebibliography}{99}

\bibitem{HarlandLang:2011qd}
  L.~A.~Harland-Lang, V.~A.~Khoze, M.~G.~Ryskin, W.~J.~Stirling,
    arXiv:1105.1626 [hep-ph]

\bibitem{KMRprosp} V.~A.~Khoze, A.~D.~Martin, M.~G.~Ryskin,
  Eur.\ Phys.\ J.\  C {\bf 23}, 311 (2002)


\bibitem{fp420} M.G.Albrow \emph{et al.}, 
J. Inst. {\bf 4} T10001 (2009)

 \bibitem{Albrow:2010zz}
  M.~Albrow,
  arXiv:1010.0625 [hep-ex]

\bibitem{Khoze04}
  V.~A.~Khoze, A.~D.~Martin, M.~G.~Ryskin, W.~J.~Stirling,
  Eur.\ Phys.\ J.\  C {\bf 35}, 211 (2004)

\bibitem{HarlandLang09}
 L.~A.~Harland-Lang, V.~A.~Khoze, M.~G.~Ryskin, W.~J.~Stirling,
  Eur.\ Phys.\ J.\  C {\bf 65}, 433 (2010)


\bibitem{HarlandLang10}
  L.~A.~Harland-Lang, V.~A.~Khoze, M.~G.~Ryskin, W.~J.~Stirling,
  Eur.\ Phys.\ J.\  {\bf C69 }, 179 (2010)

\bibitem{khrysthal}
  L.~A.~Harland-Lang, V.~A.~Khoze, M.~G.~Ryskin, W.~J.~Stirling,
  Eur.\ Phys.\ J.\  {\bf C71 }, 1545 (2011)



\bibitem{HKRSTW}S.~Heinemeyer {\it et al.}
  Eur.\ Phys.\ J.\  C {\bf 53}, 231 (2008)

\bibitem{HKRSTW1}
S.~Heinemeyer et al.
  arXiv:1012.5007 [hep-ph]

\bibitem{shuvaev}
  V.~A.~Khoze, A.~D.~Martin, M.~G.~Ryskin, A.~G.~Shuvaev,
  Eur.\ Phys.\ J.\  C {\bf 68}, 125 (2010)

\bibitem{Albrow:2010yb}
  M.~G.~Albrow, T.~D.~Coughlin, J.~R.~Forshaw,
  Prog.\ Part.\ Nucl.\ Phys.\  {\bf 65 } (2010)  149-184

\bibitem{cdf:2007na}
  T.~Aaltonen {\it et al.}  [CDF Collaboration],
  Phys.\ Rev.\ Lett.\  {\bf 99}, 242002 (2007)

\bibitem{Aaltonen:2009kg}
  T.~Aaltonen {\it et al.}  [CDF Collaboration],
  Phys.\ Rev.\ Lett.\  {\bf 102}, 242001 (2009)

\bibitem{Khoze:2004ak}
  V.~A.~Khoze, A.~D.~Martin, M.~G.~Ryskin, W.~J.~Stirling,
  Eur.\ Phys.\ J.\  C {\bf 38}, 475 (2005)

 \bibitem{Khoze00a}
  V.~A.~Khoze, A.~D.~Martin and M.~G.~Ryskin,
  Eur.\ Phys.\ J.\  C {\bf 19}, 477 (2001)

\bibitem{Brodsky81}
  S.~J.~Brodsky, G.~P.~Lepage,
  Phys.\ Rev.\  {\bf D24 } (1981)  1808

  
\bibitem{Benayoun89}
  M.~Benayoun, V.~L.~Chernyak,
  Nucl.\ Phys.\  {\bf B329 } (1990)  285

\bibitem{Brodsky0s82}
  S.~J.~Brodsky, R.~W.~Brown,
  Phys.\ Rev.\ Lett.\  {\bf 49 } (1982)  966

\bibitem{Gilman:1987ax}
  F.~J.~Gilman, R.~Kauffman,
  Phys.\ Rev.\  {\bf D36 } (1987)  2761


\bibitem{Mangano90}
  M.~L.~Mangano, S.~J.~Parke,
  Phys.\ Rept.\  {\bf 200 } (1991)  301-367


\end{thebibliography}
\end{document}